\documentstyle{elsart}
\newcommand{ \ceren}{\v{C}erenkov }
\newcommand{ \Cfour}{C$_4$F$_{10}$ }

\begin{document}
\begin{frontmatter}

\title{In-beam tests of a ring imaging \v{C}erenkov detector
       with a multianode photomultiplier readout}

\author{R. Debbe},
\author{S. Gushue},
\author{B. Moskowitz},
\author{J. Norris\thanksref{KSU}},
\author{J. Olness},
\author{F. Videb\ae k}
\address{Brookhaven National Laboratory,
             Upton, New York 11973, USA}
\thanks[KSU]{Now at Department of Physics, Kansas State
University, Manhattan, KS 66506.}

\begin{abstract}
A ring-imaging \v{C}erenkov counter read out by a 100-channel PMT
of active area 10$\times$10 cm$^2$ was operated successfully in
a test beam at the BNL AGS with several radiator gases, including
the heavy fluorocarbon \Cfour.
Ring radii were measured for electrons, muons, pions and kaons
over the particle momentum range from 2 to 12 GeV/$c$,
and a best resolution of $\sigma_r/r = 2.3\%$ was obtained.
\end{abstract}
\end{frontmatter}

\section{Motivation}

Ring Imaging CHerenkov (RICH) counters are powerful tools for achieving
particle identification over a wide range of momentum
within a single detector.  The experience of the
SLD\cite{Ashford,SLD94}
and DELPHI\cite{Delphi91,Delphi94} Collaborations in
developing broad-range detectors
has shown that $\pi$/K/p separation can be achieved up to
approximately 30 GeV/$c$ and e/$\pi$ separation can be performed up
to 6 GeV/$c$.  There is also the capability of handling modest
multiplicities within the detector.  For reviews on the subject,
see Refs.~\cite{Ypsilantis} and \cite{Leith}.

For a given particle the frequency distribution of
\ceren radiation,
$dN/d\nu$, is constant for all frequencies, $\nu$, leading to
several possible photon detection schemes.
Existing RICH counters generally detect the \ceren photons in
the UV region (6.5 -- 7.5\,eV) by means of photoionization of a complex
organic molecule, such as TMAE or TEA, in a gas chamber.
Alternatively, a segmented optical readout based on an array of
photomultipliers offers four main advantages over the gas-based
chamber:
\begin{enumerate}
\item No special handling for hazardous materials, such as
       gaseous photocathodes.
\item  No need for the high purity UV-transparent materials called for
       when detecting photons of energy greater than 6 eV
       ($\lambda < 200$ nm), such as is the
        case for gaseous photocathodes as well as CsI photocathodes.
\item  Greater overall integrated efficiency due to the wide bandwidth
       (2.5--4.3 eV) of the photocathodes used in PMT's.
\item  The possibility of readout fast enough for
       first-level triggering of the host experiment.
\end{enumerate}

For these reasons, the proposed BRAHMS Experiment\cite{Brahms} at RHIC
plans to use such a RICH counter to perform the high momentum particle
identification for a forward-angle ($2^\circ < \theta < 18^\circ$)
spectrometer.  This counter is designed to identify pions in the
momentum range from 4 to 20 GeV/$c$, kaons from 9 to 20 GeV/$c$,
and protons from 17 to 35 GeV/$c$.  To obtain the high index of
refraction necessary for the required coverage ($n$=1.00185), a
radiator of gaseous perfluorobutane (\Cfour) at 1.25 atm.~absolute
pressure has been selected.
For photon detection, the BRAHMS design calls for four
100-anode PMT's forming a two-by-two array, each with an active area
of 10$\times$10 cm$^2$, as described below.
Although arrays of single-anode PMT's\cite{NIMproc},
as well as an image intensifier coupled to a CCD readout\cite{Toru},
have been used successfully in previous RICH counters,
state-of-the-art large-area multi-anode PMT's offer the advantage of
many channels with a single common power supply and a fast, compact
readout.
For more details concerning the BRAHMS RICH counter, see
Refs.~\cite{Brahms} and \cite{memo}.

Key elements of the proposed RICH detector have been subjected to
in-beam testing using momentum-selected secondary beams at the BNL AGS,
namely:
\begin{enumerate}
\item  Use of the heavy fluorocarbon \Cfour gas radiator to provide
       a high index of refraction at near-atmospheric pressures.
\item  Use of a large-area (10$\times$10 cm$^2$) segmented multi-anode
       PMT for fast imaging of the various ring patterns.
\item  An off-axis displacement of the focal plane, obtained by a
       rotation of the focussing mirror, such that the focal plane
       detector is situated outside the radiator volume illuminated
       by the particle flux.
\end{enumerate}
Some pertinent results of these tests (RHIC R.~\& D.~Project 44)
are reported here.
Note that the multi-anode PMT used in these measurements
was the first product of a new design by the
Hamamatsu Corporation\cite{Hamamatsu}.
The expected performance criteria outlined in Ref.~\cite{Brahms},
based upon Hamamatsu's specifications, are largely borne out by the
present results.

\section{The Multi-anode PMT}

The Hamamatsu R-4549-01 is a 20-stage 100-anode PMT based
on a fine-mesh dynode structure, providing a current gain of
$3\times 10^6$ at a high voltage of HV = -2000 volts.
In this model a special grid of focusing electrodes has been
installed between the photocathode and the first dynode
which is designed to improve the spatial resolution and uniformity
as seen in the individual anode signals (i.e., pixel response).
The 10$\times$10 cm$^2$ photocathode is bialkali; the
segmented anode is a matrix array (10$\times$10 = 100) of
9.5$\times$9.5 mm$^2$ units with lemo outputs.
(The outer glass envelope of the PMT measures 12.8$\times$12.8 cm$^2$.)
With current gains from $10^6$ to $2\times 10^7$ (HV = -1800 to
-2500 volts) the outputs will
easily drive cable connections to appropriate ADC's. The high voltage is
supplied by a single high-voltage cable: a built-in R/C dropping chain
fixes the relative dynode voltages. A single output from the last dynode,
representing essentially the ``sum'' of all anode currents, is available
as a convenient trigger.

The entrance window is of boro-silicate glass with the typical
cutoff for wavelengths $\lambda < 300$ nm ($E > 4.1$ eV).
The quantum efficiency response
function is approximately Gaussian, with a peak sensitivity of 18\%,
centered at $E = 3.4$ eV and with FWHM = 1.8 eV.
(The tails, however, fall off
twice as fast as a true Gaussian.)
The integral of the PMT quantum efficiency, $\epsilon$, over the photon
energy, $E$, is $\int \epsilon(E)dE = 0.33$ eV.
The efficiency was found to be fairly uniform over the central region of
the PMT face, with a fall-off of 30-50\% at the edges and corners,
respectively.

\section{Measurements of the PMT Response}

Since the \v{C}erenkov ring-image is composed of multiple hits of
single photons, we were prompted to study the single-hit response.
For this purpose, the R-4549 PMT was first mounted in a light-tight box.
A photon source was made by collimating the light from a blue LED.
The collimator opening was positioned a few
millimeters away from the photocathode, and was such as to
illuminate only a single pixel at a time.  The LED was pulsed by a driver
whose intensity could be tuned low enough so that the collimator output
was either 0 or 1 photon. (This tuning requirement was checked
independently using an RCA 8854 PMT with excellent single-photon
resolution.)  The data acquisition was triggered by the same signal
which triggered the LED.
Figure \ref{fig:one_pe}a shows the ADC pulse height distribution for a
typical pixel.  This distribution is exponential
in shape with no distinct separation of the pedestal from a
one-photoelectron peak.  By contrast, the ADC distribution for the RCA
tube, shown in Fig.~\ref{fig:one_pe}b, shows such a separation.

The position dependence of the response is shown in
Fig.~\ref{fig:led_scan}.
The relative pulse height in each of three adjacent pixels is plotted for
a run in which the position of the LED relative to the phototube was stepped
in the $y$-direction.  (To improve the signal, the voltage driving
the LED was raised so as to produce many photoelectrons.)
The FWHM of the response for an individual pixel
is 10 mm with little cross-talk between neighboring pixels, as expected.
The pulse heights in neighboring pixels average only about
10\% of that in the hit pixel for a hit in the center.

\section{The Prototype Detector}
The prototype RICH counter is shown schematically in Fig.~\ref{fig:testbox}.
It was constructed from 1.9 cm thick aluminum plate, vacuum-welded at the
box edges, with a removable top attached by 50 1/4-20 screws and sealed
by a rubber gasket.  The inner dimensions are
127$\times$64$\times$46 cm$^3$.  The particle beam enters
via a 15 cm diameter mylar window of thickness 0.025 cm
(shown on the left in the figure)
and exits through an identical window at the output port.

A 15 cm diameter concave glass mirror\cite{EdmundSci}
whose front surface has been aluminized with a protective MgF$_2$ coating
and which has a focal length $f$ = 91.4 cm is situated at a
distance of $L$ = 114.3 cm from the beam entrance.
This mirror is rotated by an
angle $\alpha$ = 8$^{\circ}$ relative to the beam direction,
producing a ring image focussed at 2$\alpha$ = 16$^{\circ}$.
The reflectivity of the mirror is approximately 80--90\% at visible
wavelengths.
The mirror is mounted in a frame attached to an axle which can be
rotated about a vertical axis perpendicular to the beam direction.
The mirror angle is controlled by a 0.6 cm steel rod which moves a
47 cm arm rigidly fixed to the mirror frame, so as to allow a mirror
rotation over the range $-12^{\circ} < \alpha < +12^{\circ}$.
Due to practical space constraints, the position of the front face
of the phototube was at a distance of 84.5 cm from the center of the
mirror, causing images on the photodetector to be very slightly
out of focus.  In fitting experimental ring patterns we adopted an
effective focal length reduced by 3\% from the true focal length
to account for the displacement from the focal plane.

One hundred six Lemo signal cables are fed into the box using ordinary
Lemo feed-throughs epoxyed into place to maintain a vacuum seal.
The output signal from each of the PMT anodes is fed into an input
channel of a LeCroy 1882 Fastbus ADC for readout of the pulse height.

The device was aligned in the secondary test beam at the AGS,
with the trigger defined by a 2$\times$2 cm$^2$ scintillator, S1, just
before the entrance window of the test vessel, and a larger scintillator
paddle, S2, downstream of the exit window.
The initial alignment was accomplished using a laser placed on the known
beam axis; the mirror and detector angles were then set and checked
visually.

Two identical drift chambers, one 57 cm in front of and one 79 cm behind
the prototype counter, provided tracking for determining the particle
trajectory.  These two chambers are part of a set of four wire chambers
built for the tracking system of BNL Experiment 878 and used during the
1991 and 1992 heavy ion runs.  Each chamber consists of six planes with
wires oriented in three sets of directions.  For each wire orientation
an offset of half a cell size provides the information necessary to resolve
the left-right ambiguity present in these wire detectors.
The position resolution for each plane of the drift chambers has been
measured as being 150 $\mu$m.
Thus, the expected position of the center of
each RICH ring on the phototube is determined to a resolution of
approximately 500 $\mu$m in both the $x$ and $y$ directions.

\section{Radiator Gases and the Gas Handling System}

\Cfour (perfluorobutane) is a colorless, odorless, non-flamable gas
which is essentially inert.  In other respects, save for flamability,
it resembles its hydrocarbon counterpart, butane.  The most abundant
form resulting from commercial production is the n-isomer, a completely
fluorinated linear carbon chain.  Table \ref{table:gases} presents a
comparison of key properties for this and other radiator gases used
in the results presented here.
The optical transmission of a sample of the \Cfour gas used in the
measurements reported here was recorded with a UV spectrometer,
and it was found that the transmission of this gas is nearly 100\%
at wavelengths of 230-450 nm.  The DELPHI Collaboration
reports\cite{Delphi94}
complete transparency of pure \Cfour at shorter wavelengths down to
160 nm.

The gas-handling system used in filling the RICH chamber with
fluorocarbons is shown in Fig.~\ref{fig:rich_gas}.
The procedure adopted utilizes the fact that the
radiator gases Freon-12 (122) and C$_4$F$_{10}$ (238) have
molecular weights (given in
parentheses) much heavier than air (29) or argon (40).
For example, consider
the case where the RICH chamber has been initially filled with argon.
The C$_4$F$_{10}$, which is supplied by PCR, Inc.\cite{PCR},
is set to flow slowly into the bottom of the vessel, while the
output mixture
(Ar + C$_4$F$_{10}$) is sent through a cooled recovery tank
where the C$_4$F$_{10}$ is liquified
while the argon passes out through the exhaust.
The differing (by a factor of 6) molecular weights
result in a density gradient in the RICH vessel, such that
if the flow is slow enough (non-turbulent) the volume slowly becomes
enriched in C$_4$F$_{10}$.
The procedure can be continued until the supply in tank A is exhausted.
In our case we manually switched the two tanks A and B and repeated
the cycle using the C$_4$F$_{10}$ previously recovered in B.
Three such cycles yielded a gas mix of
approximately 90\% purity in C$_4$F$_{10}$.

In order to recover the C$_4$F$_{10}$ at the conclusion of the
experiment, a pump
was used to reverse the flow direction, as indicated by the dashed
lines in Fig.~\ref{fig:rich_gas}.
As the C$_4$F$_{10}$ is recovered into B, the observed pressure drop
in the chamber is compensated by adding argon to maintain a pressure
differential of $<$ 0.15 atm.

In principle, one can install a remote-controlled switching system to
actuate the controlling valves, so as to alternate the designated
function of the supply and recovery tanks.  This would allow an
extended period of automatic
cycling, leading to a nearly 100\% purity of C$_4$F$_{10}$,
or by reversing the cycle, a complete recovery of the fluorocarbon.

\section{In-Beam Test Results}

Figure \ref{fig:freon_data} presents results measured with
the beam tuned for negative 12 GeV/$c$ particles --- predominantly
pions --- and a gas mixture in the test box of Freon-12 and N$_2$.
The squares plotted in panel (a) are proportional in size to
the pulse height for each phototube pixel, summed over
approximately 1100 events.
The drawn ring is the best fit of the data to a circle.
Figure \ref{fig:freon_data}b is a histogram of the ADC pulse height
in pixel (3,8), which lies directly on the ring.
Note the similarity to Fig.~\ref{fig:one_pe}a.
The corresponding histogram from neighboring pixel (3,9), which is
not on the ring, is shown in Fig.~\ref{fig:freon_data}c; clearly the
cross-talk between adjacent pixels is small, although not completely
negligible.

The ability of the counter to identify individual particles is
demonstrated in Fig.~\ref{fig:rich_3gev}.  Panels a, b and c
show individual events from a run where the beam momentum was
tuned to 3 GeV/$c$ and the box contained a C$_4$F$_{10}$-Ar mixture.
The drawn rings show the best fits (see below)
to circles constrained by the
positions of the ring center (shown as the small star symbols)
given independently by the drift chamber tracking.
The ring radius from all events in the run is histogrammed in
Fig.~\ref{fig:rich_3gev}d where three distinct peaks, identified with
$\pi^-$, $\mu^-$ and e$^-$, are clearly observed.  The peaks are
well described as Gaussian in shape; for the rings which are
fully saturated in radius (electrons) the resolution of the radius
determination is $\sigma_r/r = 2.3\%$.

The fitting procedure extracted the ring radius by means of the
ADC-weighted average of the distances from pixel centers to the
ring center given by the tracking.  Exhaustive simulations of
this ring-fitting algorithm show negligible bias for rings of
radius from 2.0 to 4.2 cm, and predict a resolution of approximately
half of that actually achieved.  The remaining contribution to
the measured resolution can be attributed to chromatic and geometric
aberrations (see discussion in Ref.~\cite{Brahms}).

\section{Momentum Scan}

A momentum scan was performed with a C$_4$F$_{10}$-Ar mixture in the
vessel during a period in which the mixing proportion of the two gases
was held fixed.
The ring radius is plotted versus momentum
in Fig.~\ref{fig:momentum_scan}, which shows saturated light production
from electrons, as well as the excitation curves for muons, pions
and kaons.
The saturated radius for electrons implies an index of refraction
$n$=1.00113, shown as the horizontal line through the electron points.
The excitation curves predicted by this index of refraction are shown
as the solid curves; the shaded regions represent a $\pm$5\%
systematic uncertainty in the absolute momentum scale.  The measured
values of the radius all lie within the shaded regions, giving
confidence that the particle identification is correct.
{}From $n$=1.00113 it is inferred that the fraction of C$_4$F$_{10}$ in
the vessel was approximately 75\%.  Later runs following further
filling of the vessel achieved approximately 90\% C$_4$F$_{10}$.

\section{Photoelectron Yield and the Figure of Merit}

The difficulty in determining the photoelectron yield precisely can
be understood from examination of the single-pixel response functions
illustrated in Fig.~\ref{fig:freon_data}b and c.  The wire-mesh dynode
structure of the H-4549 PMT does not provide the typical Gaussian
response function which would allow one to distinguish 1, 2 or 3
photoelectron hits.  Rather, the exponential shape of the response
function makes it difficult to determine for a given event
the number of hits on an individual pixel.

The yield was instead deduced from the distribution of the
anode ADC sum
(pedestal and gain corrected) per event.  For the case of the
\Cfour-Ar mix mentioned in the previous section and 12 GeV/$c$
pions, the ADC sum distribution perfectly describes a gaussian
curve whose mean is 699 and r.m.s. width is 185; hence the number
of photoelectrons is (699/185)$^2$=14.3, and the figure of merit
is approximately 60 cm$^{-1}$.
Although this value
is somewhat lower than the expected value of 90 cm$^{-1}$, as based
on the manufacturer's specifications for the mirror and PMT,
it suffices for the design criteria of BRAHMS\cite{Brahms}.

\section{Conclusions}

All three key elements of the BRAHMS RICH counter, namely the use of
\Cfour gas, the use of the large area multi-anode PMT, and the off-axis
displacement of the focal plane, were successfully tested together.
Particle identification was possible on an event-by-event basis
by determining the \v{C}erenkov ring radius to a resolution of
$\sigma_r/r = 2.3\%$,
which is quite close to the design value in Ref.~\cite{Brahms}.
Improvement in
the systematic understanding of the ring radius versus
momentum curves (see Fig.~\ref{fig:momentum_scan}) is planned for
future measurements which will include determination of the particle
momentum by time-of-flight techniques and direct measurement of the
radiator index of refraction using a Fabry-Perot interferometer.

\ack

We would like to thank E. Hergert and S. Suzuki of Hamamatsu Corp.,
as well as K. Arisaka of UCLA, for useful discussions and developments
on the special multi-anode PMT used in this work.
We also thank our colleagues, C.~Chasman, H.~Hamagaki and T.~Sugitate,
for their interest and assistance in the initiation of this
investigation.
Consultations with R. Du Boisson, of PCR Inc., on the nature of the
fluorocarbon gases is also gratefully acknowledged.
We also want to thank R.A. Holroyd of the BNL Chemistry Department
for his measurement of the UV transmission of \Cfour.
This work was supported by
U.S. Department of Energy contract number DE-AC02-76CH00016,
in part through the R.\&D. funds of the RHIC Project,
and we thank T.~Ludlam for his encouragement in this enterprise.

\newpage
\begin{table}[p]
\begin{tabular}{|l|r|r|r|c|r|}
\hline
   & b.p.        & $M$~~~ & $C_p~~~~$  & v.p. &  $(n-1)\times10^6$ \\
   & ($^\circ$C) & (g/mol) & (Joule/mol$\cdot$K) & (atm.) & \\
 \hline
 N$_2$           & -196 &   28.1~ &  29.12~ & ---   &  285~~~ \\
 Ar              & -186 &   39.9~ &  20.79~ & ---   &  269~~~ \\
 Freon-12        &  -30 &  120.9~ &  73.35~ & 5.77  & 1180~~~ \\
 C$_4$F$_{10}$   &   -2 &  238.0~ & 243.56~ & 3.24  & 1415~~~ \\
\hline
\end{tabular}
\caption[Properties of Radiator Gases]
{Properties of radiator gases used in the results of this paper.
Listed are the boiling point (b.p.) in degrees Celsius, the molecular
weight
($M$), the constant-pressure heat capacity ($C_p$) at 300 K,
the vapor pressure (v.p.) in atmospheres,
and ($n$-1)$\times$10$^6$, where $n$ is the
index of refraction at 1 atm.~pressure,
300 K and a wavelength of 300 nm ($E$=4.1 eV).
The value of $C_p$ for \Cfour was calculated using the method of
Benson (Ref.~\cite{Benson}) and the tables in Ref.~\cite{Reid}.
The index of refraction for Freon-12 is deduced from the data
given in Ref.~\cite{Hayes}.
The index of refraction for \Cfour
was calculated using the Lorentz-Lorenz Equation and measurements
reported in Ref.~\cite{Ypsilantis}.}
\label{table:gases}
\end{table}

\newpage
{\large Figure Captions}

\begin{figure}[h]
\caption{Single photoelectron ADC distributions from LED measurements
   a) for the Hamamatsu multi-anode PMT  b) for an RCA 8854 PMT.}
\label{fig:one_pe}
\end{figure}

\begin{figure}[h]
\caption{Position dependence of the PMT pulse height in three adjacent
         pixels, measured as the position of an LED is stepped with
	 respect to the phototube.}
\label{fig:led_scan}
\end{figure}

\begin{figure}[h]
\caption{Schematic view of the prototype RICH counter.
         See text for details.~~~~~~~~~~}
\label{fig:testbox}
\end{figure}

\begin{figure}[h]
\caption{Schematic flow-diagram for gas-handling system.
         Solid lines are for initial filling with the fluorocarbon;
         dashed lines (note reversed direction) are for
         final recovery of the fluorocarbon gas. With this system,
         the pressure in the RICH vessel can be kept within $<$ 10\%
         of atmospheric over both the filling and recovery operations.}
\label{fig:rich_gas}
\end{figure}

\begin{figure}[h]
\caption{Results from the prototype RICH counter
         with test beam at 12 GeV/$c$ and a Freon gas mixture.
	 Panel a shows a ring image
	 from approximately 1100 events and panels b and c show
	 the ADC pulse height spectra from two individual pixels.
         See text for details.}
\label{fig:freon_data}
\end{figure}

\begin{figure}[h]
\caption{Results from the prototype RICH counter
         with test beam at 3 GeV/$c$ and a C$_4$F$_{10}$-Ar gas
	 mixture.  Panels a, b and c show
	 individual events with fitted rings; the ring centers
	 are shown as stars and come from independent drift chamber
	 tracking; the box sizes are proportional to the pulse
	 height in each pixel.
	 Panel d is a histogram of the fitted ring radius
	 from the entire run, showing peaks identified as $\pi^-$,
	 $\mu^-$ and e$^-$.}
\label{fig:rich_3gev}
\end{figure}

\begin{figure}[h]
\caption{The mean radius from event-by-event fits to rings in the
	 prototype RICH counter filled with a C$_4$F$_{10}$-Ar gas
	 mixture, for different particle species.  The solid curves
	 show the expected radii for an index of refraction of
	 $n=1.00113$; the shaded regions represent a 5\% systematic
	 uncertainty in the absolute momentum scale.}
\label{fig:momentum_scan}
\end{figure}

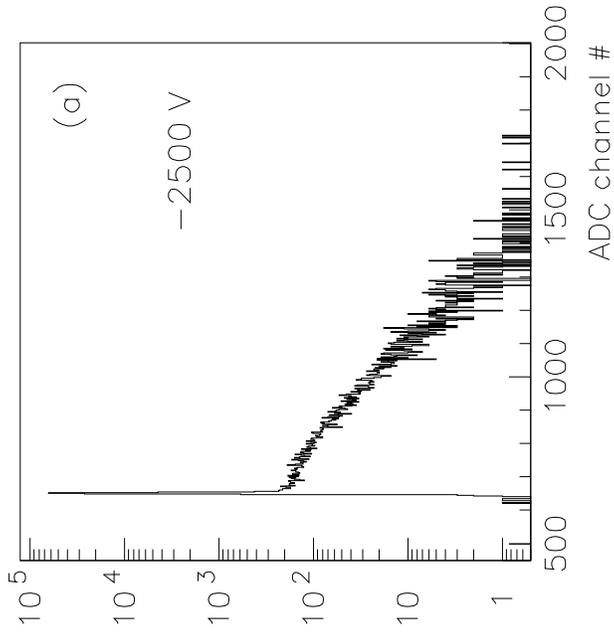
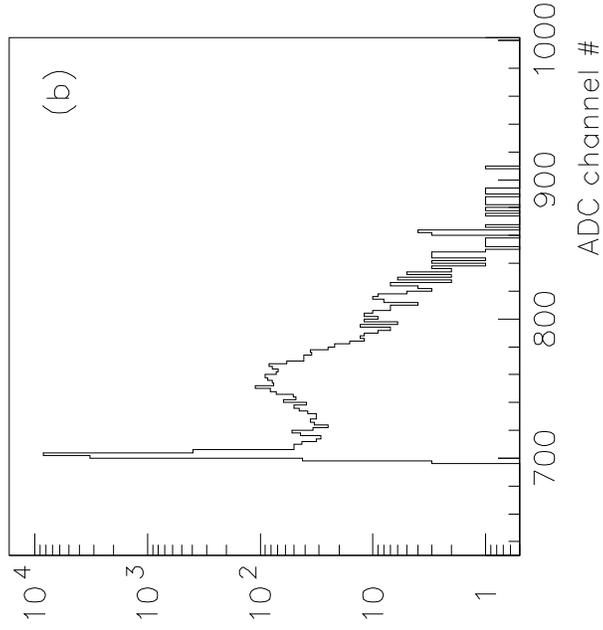
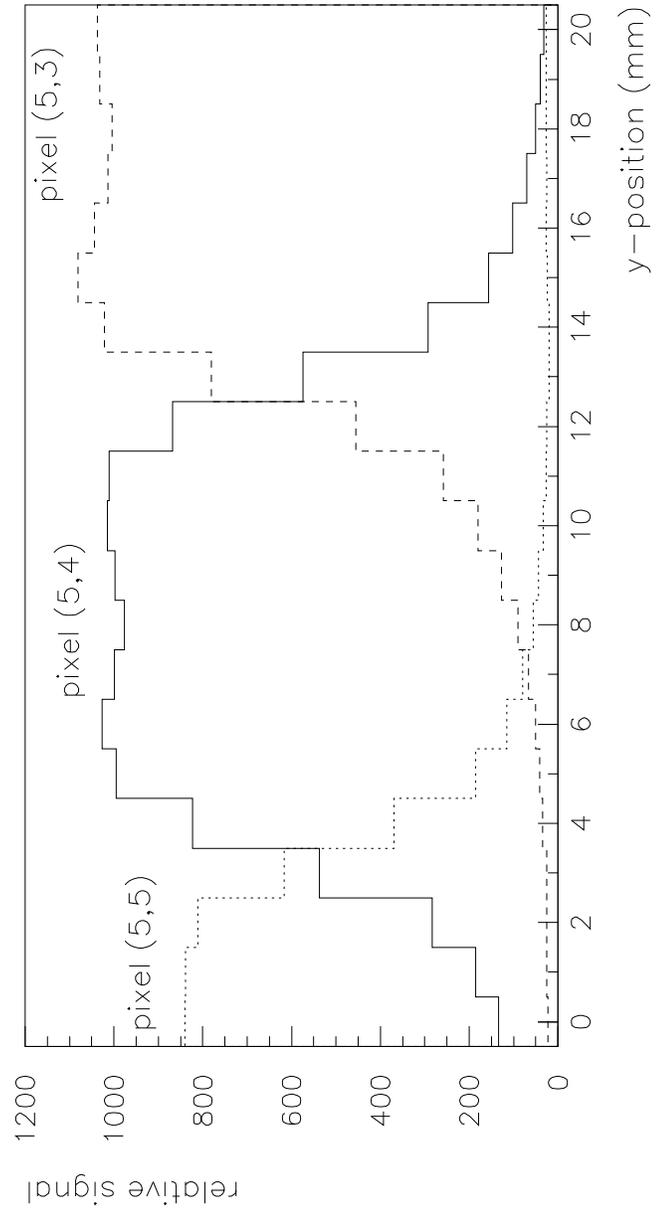
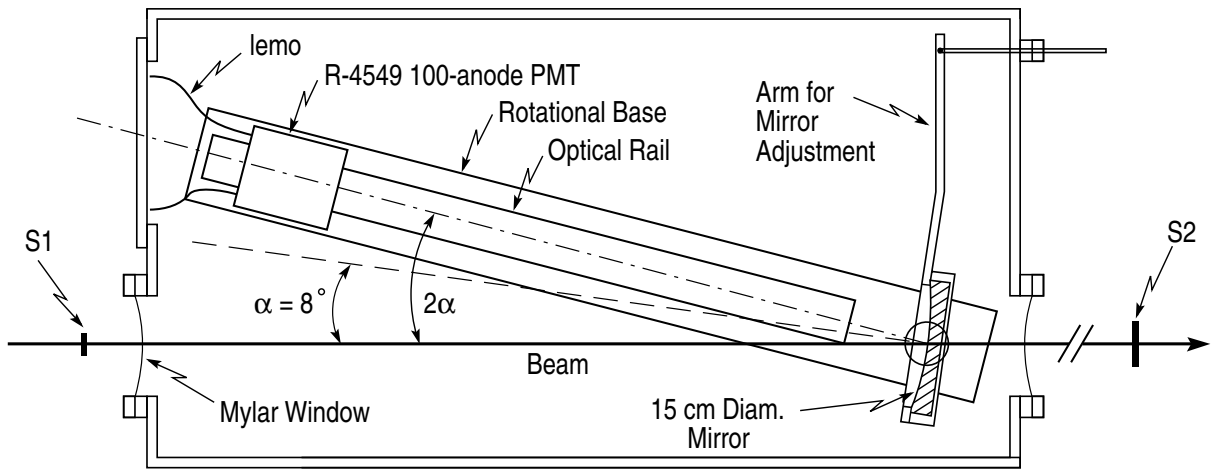
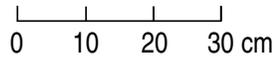
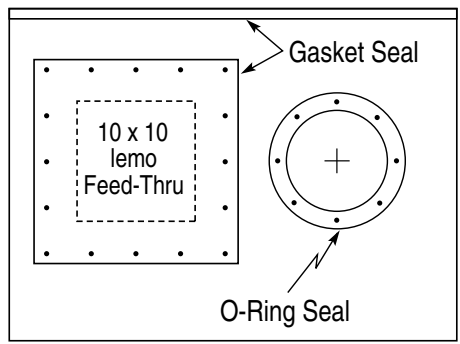
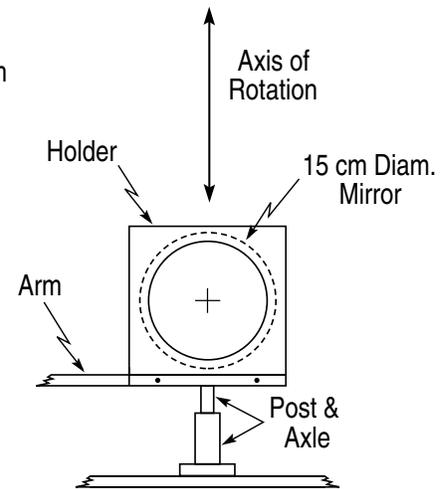
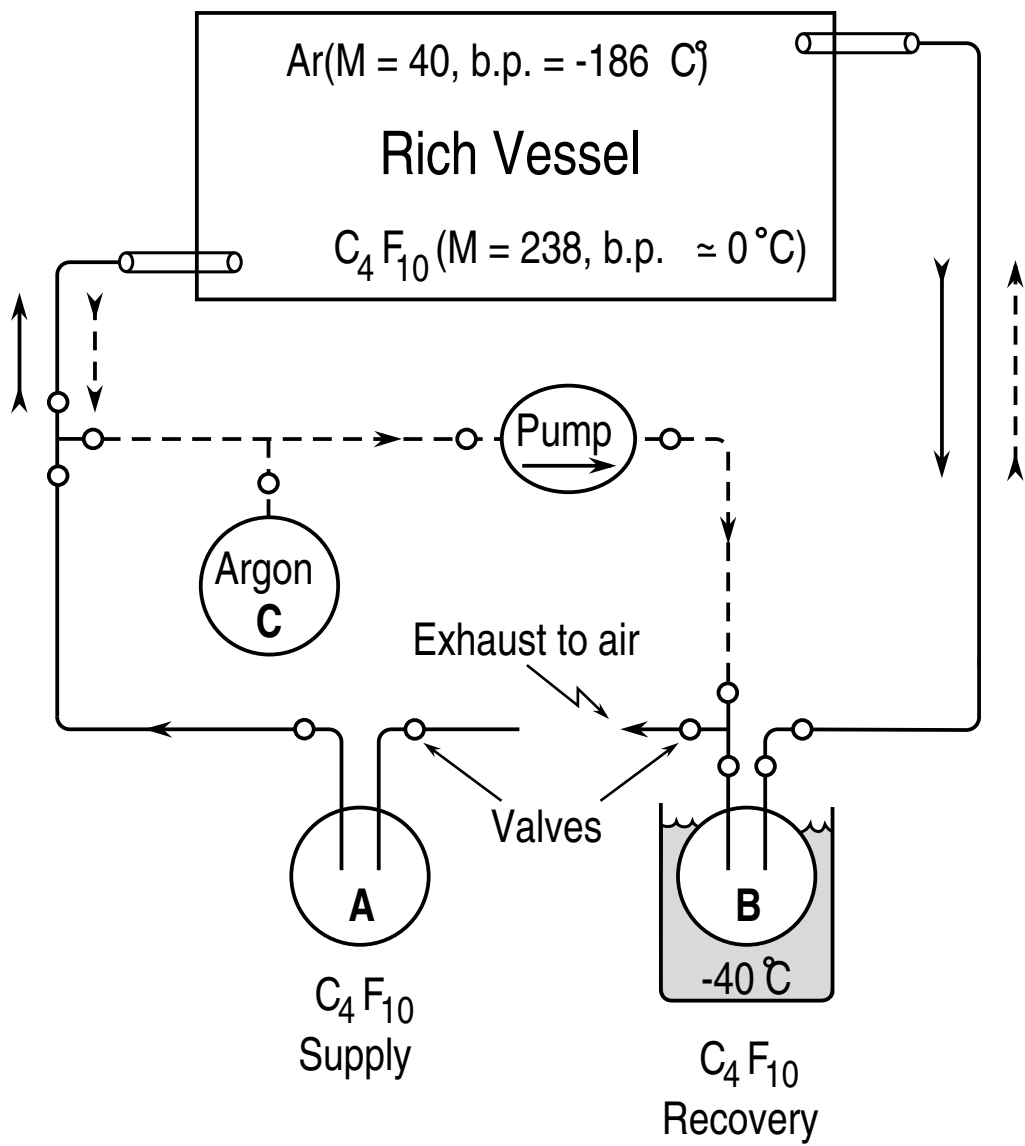
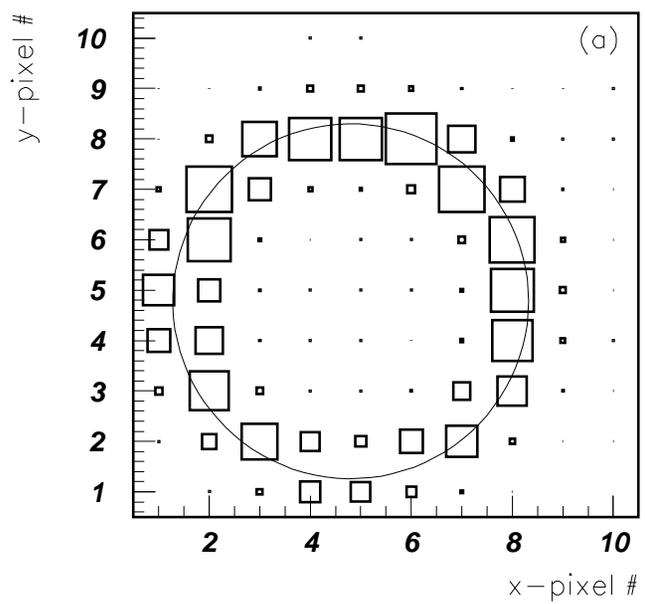
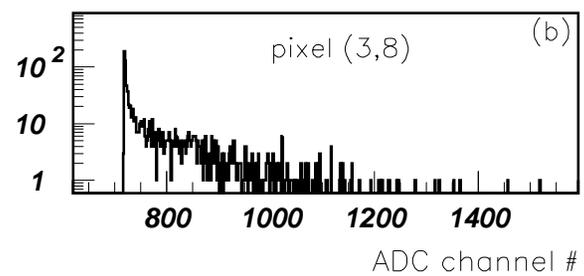
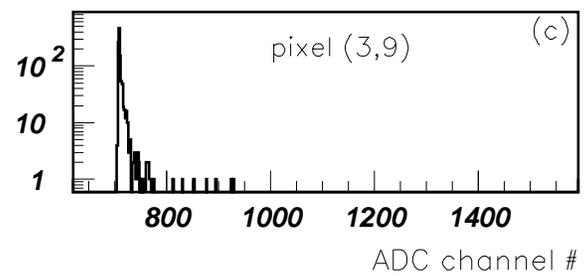
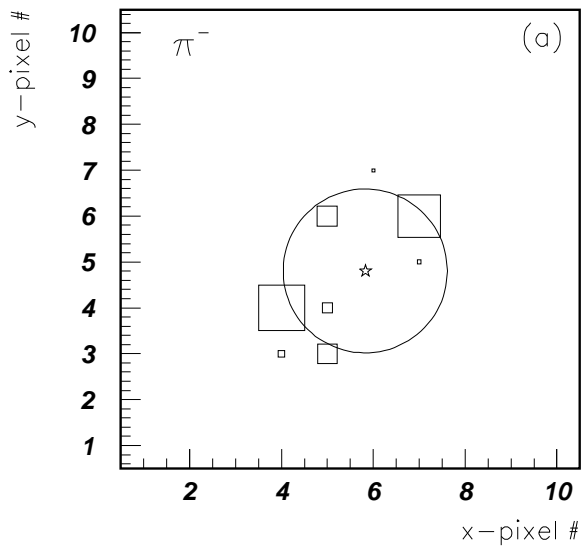
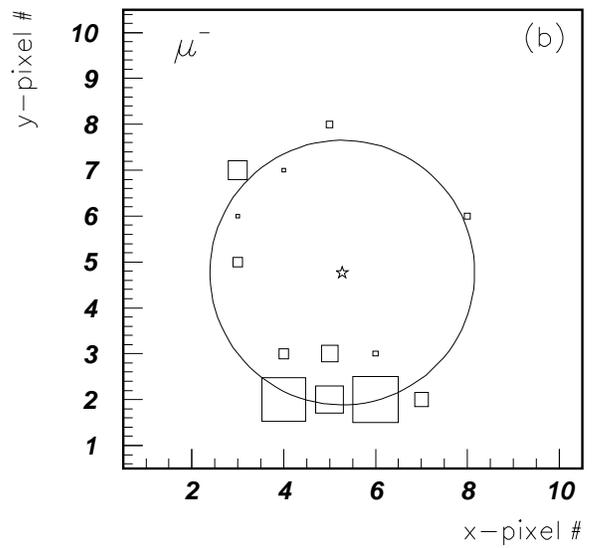
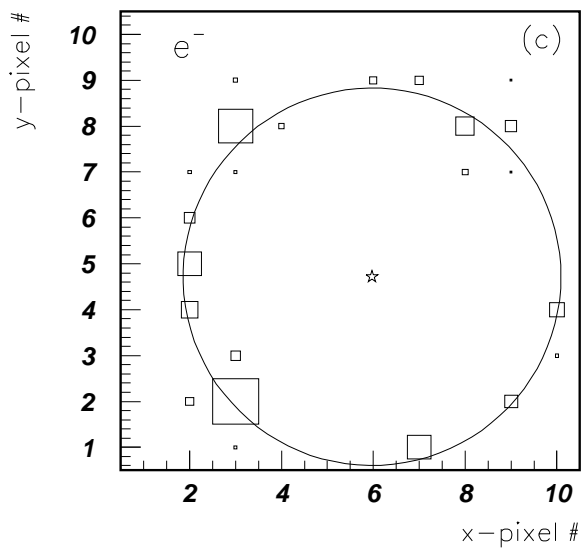
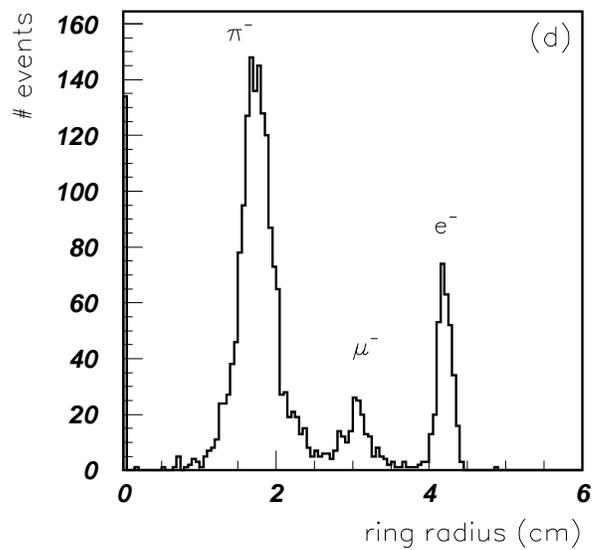
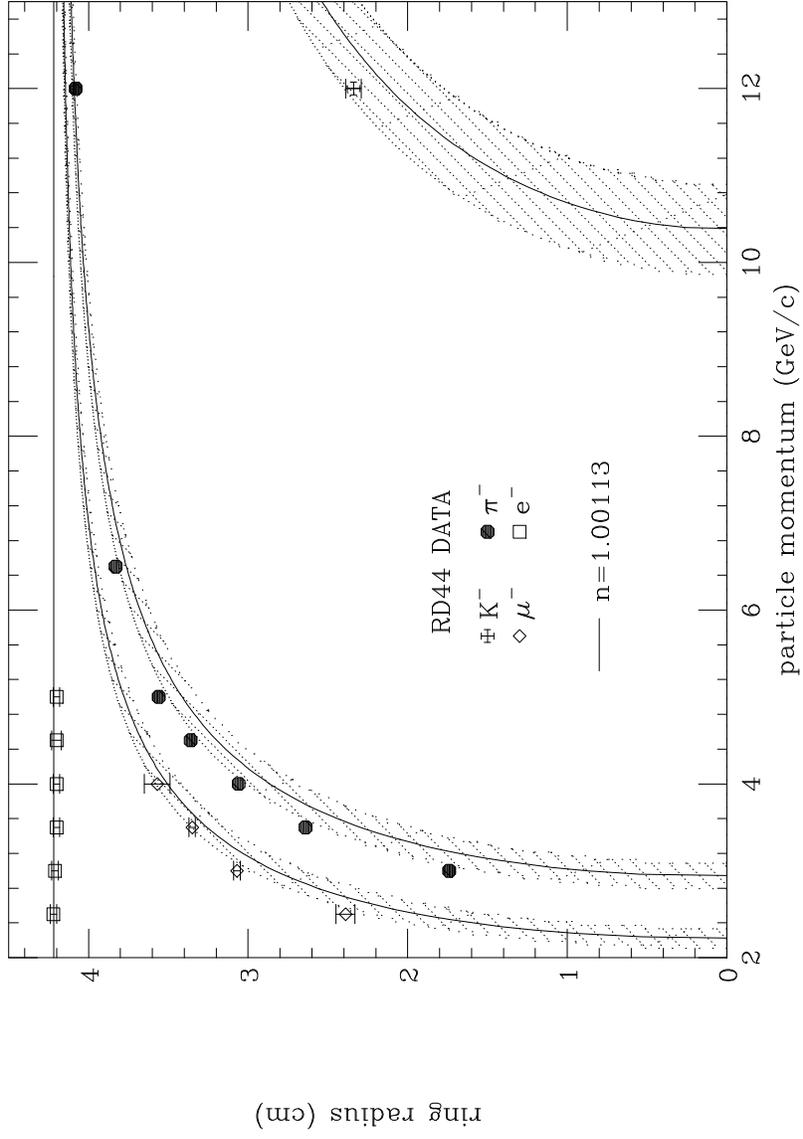
\end{document}